\documentstyle[12pt,aaspp4]{article} 
\begin{document}

\title{IRAS~06562$-$0337, The Iron Clad Nebula: \\
A New Young Star Cluster}

\author{David R. Alves$^{1}$ }
\affil{Lawrence Livermore National Laboratory, Livermore, CA 94550\\
E-mail:  alves@igpp.llnl.gov}

\altaffiltext{1}{Department of Physics, University of California, Davis, CA 95616} 

\author{D. W. Hoard}
\affil{Department of Astronomy, University of Washington,
Seattle, WA 98195 \\
E-mail: hoard@astro.washington.edu}

\author{Bernadette Rodgers}
\affil{Department of Astronomy, University of Washington,
Seattle, WA 98195 \\
E-mail:  rodgers@astro.washington.edu}

\begin{abstract}

IRAS~06562$-$0337 has been the recent subject of a classic debate:
proto-planetary nebula or young stellar object?
We present the first 2$\micron$ image of IRAS~06562$-$0337, which
reveals an extended diffuse nebula containing approximately 70 stars 
inside a 30$\arcsec$ radius around a bright, possibly resolved, central
object.  The derived stellar luminosity function is consistent with that
expected from a single coeval population, and the brightness of the
nebulosity is consistent with the predicted flux of unresolved low-mass stars.
The stars and nebulosity are spatially coincident with strong CO line emission.
We therefore identify IRAS~06562$-$0337 as a new young star cluster embedded
in its placental molecular cloud.  The central object is likely a Herbig Be
star, $M \approx 20 M_{\odot}$, which may be seen in reflection.  We
present medium resolution, high S/N, 1997 epoch optical spectra of the
central object.  Comparison with previously published spectra shows new
evidence for time variable permitted and forbidden line emission, including
\ion{Si}{2}, \ion{Fe}{2}, [\ion{Fe}{2}], and [\ion{O}{1}].  We suggest
the origin is a dynamic stellar wind in the extended, stratified atmosphere
of the massive central star in IRAS~06562$-$0337.

\end{abstract}

\clearpage

\section{Introduction}

Garcia-Lario, Manchado, Sahu, and Pottasch (1993, hereafter GMSP)
present the first detailed analysis
of IRAS~06562$-$0337.  They argue that it is
a proto-planetary nebula (PPN) undergoing final mass-loss
episodes.  Their time-series of optical spectra, obtained over
a 5 year period, show the onset of forbidden line emission and
the possible evolution of the central star toward hotter temperatures.
They derive a Zanstra temperature of
2$\times$10$^{4}$ K, with a slight increase over a two year
interval.  The effective temperature
of the exciting star, T$_{eff}$ $\approx$ 3.6$\times$10$^{4}$ K, also
showed a slight increase in two years.
The H$\alpha$ line profile changes in time, which GMSP interpret
as variable high velocity winds associated with episodic mass-loss.
The appearance of [\ion{O}{3}] emission lines
in 1990 and the resulting 4363/(4959+5007)
line ratio requires an ionizing region of
high electron density, log($n_{e}$) $\approx$ 6.9. 
The absence of these lines in spectra obtained before and
after 1990 is interpreted as collisional de-excitation
due to changing densities in the ionized region effected
by violent episodic mass-loss.
From CO observations GMSP derive $V_{LSR}$ = 50 $\pm$ 1 km sec$^{-1}$, 
which agrees with the velocity derived from their high resolution optical spectra.
Adopting a model galactic rotation curve, they estimate a distance
of 4 kpc, which compares to a distance of 2.4 kpc estimated from
the equivalent width of Na D absorption seen in their spectra.
The IRAS colors fit
with blackbodies show a trend of decreasing temperature with increasing
wavelength which implies a gradient of dust temperatures.  GMSP
integrated the optical--IR spectral energy distribution of
IRAS~06562$-$0337, yielding a luminosity
of L = 7000 $L_{\odot}$ for their preferred distance of 4 kpc.  

Kerber, Lercher, and Roth (1996, hereafter KLR) describe 
an additional medium resolution, high S/N spectrum of IRAS~06562$-$0337
obtained in early 1996.  [\ion{O}{3}] emission is still absent,
but a wealth of \ion{Fe}{2} and [\ion{Fe}{2}] lines are found.  These lines
confirm the high electron density derived by GMSP from the [\ion{O}{3}]
lines present in 1990.  KLR argue that the spectrum also implies
a considerable density gradient in the object, as [\ion{Fe}{2}] lines
are collisionally suppressed at densities where \ion{Fe}{2}
lines exist.  They maintain the classification
of IRAS~06562$-$0337 as a candidate PPN, designating it
``The Iron Clad Nebula''.

Bachiller, Gutierrez, and Garcia-Lario (1998, 
hereafter BGG) present new mm and sub-mm observations
of IRAS~06562$-$0337.
They derive $V_{LSR}$ = 54.0 $\pm$ 0.2 km sec$^{-1}$ and
adopt a different model Galactic rotation curve than
GMSP to estimate a distance of 7 kpc.
This distance yields a luminosity
of 21000 $L_{\odot}$ and a cloud mass M $>$ 1000 $M_{\odot}$.
From the strength of the CO emission
and the presence of CS emission, BGG surmise that IRAS~06562$-$0337
is a ``young stellar object (or small cluster) still associated
to its parent molecular cloud.''
BGG point out that IRAS~06562$-$0337 satisfies the three criteria
for a Herbig Ae/Be star (Herbig, 1960) and the spectral energy distribution,
which rises sharply in the far infrared (GMSP), is similar to Group II
Herbig Ae/Be stars (Hillenbrand, 1992).
They also note the presence of blue and
redshifted wings in the CO emission indicating a bipolar outflow.
The CO outflow may be driven by an eruptive ionized jet, which
leads them to suggest the sporadic [\ion{O}{3}] emission seen by
GMSP originates in a Herbig-Haro object.

We present the first 2$\micron$ image of IRAS~06562$-$0337.
Our image reveals a compact cluster of 
stars surrounding a bright, central object.  We independently
confirm the result also discovered by BGG that IRAS~06562$-$0337 
is a young stellar object.  
In Section 2 of this paper,
we describe our near-infrared observations and 
stellar census of the IRAS~06562$-$0337 cluster.  We also
compare the CO(2$\rightarrow$1) map of BGG with our image. 
In Section 3, we describe our new spectroscopic
observations and summarize the resulting 1997 epoch 
emission line data.  We make a detailed comparison
with the 1996 epoch emission line data of KLR.
In Section 4, we present our conclusions.

\section{Near-Infrared Imaging}

\subsection{Observations and Data Reduction}

On 1997 March 26 UT, we observed IRAS~06562$-$0337 with a
K$^{\prime}$ (1.95 to 2.35 $\micron$) filter in non-photometric
conditions using the Lick Observatory 3m telescope and the
Lirc~II mercury-cadmium-telluride 256$\times$256 pixel
camera (Misch, Gilmore, and Rank 1995).  
The wide field-of-view optical configuration
was utilized to yield a pixel size of 0.57$\arcsec$ and
a full image area covering 2.43$\arcmin\times$2.43$\arcmin$.
We implemented a four-point, on-source dithering pattern
to obtain 5$\times$2-second exposures at each position; 
this pattern was repeated 10 times.  Our cumulative
exposure was 400 seconds.
Evening twilight sky flat and morning dark 
calibration frames were obtained on the same night.
All reductions were done with IRAF\footnote{The Image Reduction and Analysis
Facility, v2.10.2, operated by the National Optical Astronomy Observatories.}.
The 200 object images were dark subtracted and flat corrected (bad pixels were
also masked), then individually sky subtracted, registered, and combined.  The field 
of view exposed for the complete 400 seconds was 108$\arcsec$~$\times$~108$\arcsec$,
and the final image was trimmed to this size.  

A log-scaled greyscale image of IRAS~06562$-$0337 with a
field of view of 70$\arcsec$~$\times$~70$\arcsec$ is presented
in Figure 1.  
The image reveals a small, dense cluster of stars around a bright, central
object.  The association of the variable emission-line central object  
with a cluster of stars lends strong support to its classification as
a young stellar object and not a PPN.  Also evident is a diffuse
nebulosity extending approximately 30$\arcsec$, whose brightness
increases toward the central object.
This nebulosity may be reflected light from the central object or
the unresolved light from numerous low-mass stars in the cluster.  
The morphology of
the faint nebulosity is somewhat affected by our choice of an
on-source dithering pattern and the resulting sky subtraction.  In
particular, directly to the East of the central object, the apparent
decrease in the brightness of the nebulosity is an artifact
of the sky subtraction.  The log-scaling emphasizes this low level data
reduction artifact, which does not affect the results of this paper. 
An improved observing strategy for IRAS~06562$-$0337 would require
chopping well away from the cluster to obtain sky images.

Our final image was photometered with the stand-alone
DaophotII/AllstarII (Stetson, 1987).  An empirical PSF
was derived from 7 bright, fairly isolated stars with profile
errors of order 3\%.  We obtained satisfactory star subtraction
and PSF fitting photometry for all cluster stars except the central object.
The seeing in our image measured with the PSF stars is 1.3$\arcsec$ (FWHM),
while the central object has a FWHM of 1.5$\arcsec$.  Thus we may be just 
resolving the central object.
As conditions were not photometric, we can only crudely
calibrate our photometry to K$_{CIT}$.  Manchado
et al. (1989) give K = 9.15 $\pm$ 0.02 mag for IRAS~06562$-$0337
obtained with a photometer and a beam width of 15$\arcsec$.  This
K magnitude is on the Tiede photometric system which is equivalent
to K$_{CIT}$ (Arribus and Martinez, 1987).
We derived an aperture correction using a 15$\arcsec$ beam
centered on the central object and the reported AllstarII magnitude.
Given the slightly non-stellar profile of the central object, this
procedure yields a rather large uncertainty of $\pm$ 0.2 mag
for the zero-point transformation from our instrumental
system to K$_{CIT}$.  With the exception of the central
object, our relative photometry of the cluster stars
is more accurate.  Typical AllstarII reported errors
are K = 12.5 $\pm$ 0.03 mag, K = 16.5 $\pm$ 0.1 mag, and
K = 18 $\pm$ 0.5 mag (see Fig.~2).

Artificial star tests were performed with DaophotII in 10 trials
of adding 50 stars to our K$^{\prime}$ image, randomly distributed across
the image and with magnitudes from K = 14 to 20.  We find our completeness,
defined as the ratio of stars recovered to stars added, is dependent
on both magnitude and position.  Considering 
artificial stars added at all positions in the image, we
are approximately 95\% complete at magnitudes brighter than
K = 15, 90\% complete at K = 16, and 50\% complete at K = 17.  
At K = 18 we recovered
almost no added stars.  We find that within a radius of $\approx$ 5$\arcsec$ 
from the central object, we recovered no stars.  Considering only
artificial stars with K $<$ 17, we find our completeness to 
increase from approximately 85\% around 10$\arcsec$ from
the central object to nearly 100\% at a radial 
position 60$\arcsec$ from the central object.  This radial
dependence of our completeness is attributable to both crowding
and the irregular sky background contribution of the nebulosity.

\subsection{ Stellar Census }

Our current near-infrared data is not ideal
for a stellar census of IRAS~06562$-$0337.  We primarily
require a wider field of view (e.g. a mosaic image) to better characterize the 
foreground (and background) stars along this line of sight
through the galactic plane.
Nevertheless, as a first accounting of the cluster members we 
adopt a radius of 30$\arcsec$ as the cluster boundary, which
is roughly the extent of the nebulosity seen in our
K$^{\prime}$ image.
Inside this radius we find 71 stars.  Two equal area control
fields, rectangular regions at the outer-most edges of our final image, 
contain 9 and 17 stars,
predicting an average of only 13 background/foreground stars will be found
inside our 30$\arcsec$ radius cluster boundary.
We conclude that there is a significant
stellar surface density enhancement in the immediate vicinity
of IRAS~06562$-$0337.  This confirms what may seem obvious
from a casual inspection of Figure 1, that we have discovered a new 
star cluster.  

In Figure 2, the top panel, we present the K luminosity function 
of stars found inside our 30$\arcsec$ radius cluster boundary.  The cluster
luminosity function is shown as the unshaded solid line histogram.  
The averaged luminosity
function of our two control fields is shown as the shaded histogram.
These two luminosity functions sample the same area of sky.  Except
for small differences in photometric completeness near the cluster center, 
they are properly comparable.
A K-S test on the cluster and background luminosity functions gives
a 1\% probability that these are drawn from the same distribution.
Corrections for completeness near the cluster center would result
in even lower probabilities.
Thus, in addition to the clustering in 2 dimensions on the sky,
we find the K magnitude distribution of stars centered on 
IRAS~06562$-$0337 is not likely to be
a statistical fluctuation in starcounts along this line-of-sight 
through the galactic plane.

In the bottom panel of Figure 2, we present the cluster luminosity function
corrected for foreground/background stars and completeness.
This luminosity function has 80 stars with K $<$ 17.5 mag.
We have additionally overplotted theoretical luminosity functions\footnote{
To calculate a theoretical luminosity function from an IMF requires a
K mag/stellar mass calibration. This is discussed in \S2.3 of this paper.}
corresponding to different initial mass functions (IMFs), parameterized by
a power-law index $\alpha$.  Our fiducial IMF is that of Salpeter (1955; 
$\alpha$ = 2.35), plotted as a solid line for K $<$ 17.5 mag and 
a long-dashed line for K $>$ 17.5 mag.
A shallower and steeper IMF ($\alpha$ = 1.35 and 3.35, respectively) are also 
shown, as dotted lines.  All three of the theoretical luminosity functions are 
normalized to have 80 stars for K $<$ 17.5 mag.
We note that the steep IMF is the best fit, although without color information
and comparisons to pre-main sequence isochrones, this analysis is susceptible to
systematic biases.  For instance, pre-main sequence stars will generally appear
brighter at K than main sequence stars, biasing the IMF slope to higher values.
We emphasize that our simple model luminosity functions demonstrate
only a plausible, order-of-magnitude consistency  between the observed cluster
luminosity function and that expected from a Salpeter IMF.
Nevertheless, this rough agreement supports the
principal result of our paper, that IRAS~06562$-$0337 is a coeval
young star cluster.

The theoretical luminosity functions allow us to estimate the 
K flux for an unresolved population of low-mass cluster stars,
perhaps seen as the diffuse nebulosity in our image.
The model luminosity functions, in this instance, can be
regarded as empirical extensions of the observed luminosity
function, and assigned no greater physical significance.
Masking the central star (using IRAF) in an image with all other identified stars
subtracted (using AllstarII), we find K$_{neb}$ $\approx$ 11.3 mag 
inside a 30 $\arcsec$ beam.   
The measured flux of the nebulosity is quite sensitive to the adopted
procedure of masking the central star.  From several trials we
estimate an uncertainty of roughly $\pm$ 0.5 mag.  
Excluding the central star (K = 9.2) the integrated magnitude of the 
stars in the histogram of the bottom panel of Figure 2 is K = 11.0 mag;
including the central star we find K = 9.0 mag.  
Integrating the Salpeter IMF luminosity function 
beyond the last reliable bin in
the completeness corrected luminosity function, 
from K = 17.5 to 23, gives K$_{neb,lf}$ = 12.3 mag.
Integrating the steep IMF from K = 17.5 to 23 gives 
K$_{neb,lf}$ = 11.5 mag.  
We find the differences in K$_{neb,lf}$ to be small when integrating to
faint magnitude limits, changing only 0.2 mag when integrating to K = 21 or 23.
Both of the predicted nebular fluxes are fainter than the measured flux, albeit
with somewhat large uncertainties.  
Therefore, we might expect some contribution of reflected 
light from the central object to the nebulosity.  Importantly, there is no
gross inconsistency in our calculations, such as a prediction of too much 
diffuse light from faint stars.
We conclude that our theoretical fit to the K luminosity function,
extended to low mass stars (M $\approx$ 0.3 $M_{\odot}$), predicts an
unresolved flux that is consistent with the brightness of 
the nebulosity observed in our K$^{\prime}$ image.

\subsection{ Mass and Size of IRAS~06562$-$0337 }

We adopt $K = -8\log(M/M_{\odot}) + 19.4$ for stars in the IRAS~06562$-$0337 
cluster\footnote{Assuming constant values for extinction, distance modulus, bolometric
corrections, and the bolometric magnitude of the Sun, standard definitions give, 
$K \propto M_K \propto  M_{Bol} \propto -2.5\log(L/L_{\odot})$.
We adopt $\log(L/L_{\odot}) \propto 3.2\log(M/M_{\odot})$ from
Mihalas and Binney (1981, p.~113).}.
The zero-point is estimated as follows.
From Carpenter et al.~(1997, see their Fig.~6), who compared near-IR data of the young
star cluster Mon~R2 to the pre-main sequence stellar 
evolution tracks of D'Antona and Mazzitelli (1994), we estimate an 
unreddened $M \approx 2.5 M_{\odot}$ star would have K = 10.2 mag. 
Assuming a relative distance modulus between IRAS~06562$-$0337 and Mon~R2
of 5 mag, and an IRAS~06562$-$0337 extinction of $A_K$ = 1 mag, yields the 
19.4 mag zero-point. 
Our K mag/stellar mass calibration
equates the bright central object, K~=~9.2 mag, with an $M = 18.8 M_{\odot}$ star. 
Our mass estimate for the central star is consistent with an early B
or late O spectral type (Mihalas and Binney, 1981).  BGG estimate a B0-B2
spectral type from the luminosity, 21000 $L_{\odot}$.  For comparison, GMSP
calculate T$_{eff}$ $\approx$ 3.6$\times$10$^{4}$ K for the central star, which
implies a late O spectral type (Mihalas and Binney, 1981).

Using the Salpeter IMF theoretical luminosity function to extend our
observed luminosity function, we 
predict the following total numbers of cluster stars: 225, 445, 875, 1720, 3380
to the limiting magnitudes K = 19, 20, 21, 22, and 23 respectively.  
The total mass in the observed luminosity function is estimated by integrating
the Salpeter IMF theoretical luminosity function to K = 17.5 which gives
$M_{stars}$ $\approx$ 225 $M_{\odot}$.  If we integrate 
to K = 23, we find $M_{stars}$ $\approx$ 1950 $M_{\odot}$, which compares
with the lower limit of molecular gas mass given by BGG of 
$M_{gas} > 1000 M_{\odot}$.
If a 50\% mass-efficiency for star formation is adopted, we might expect
the total mass of molecular gas in the cluster to be 
$M_{gas}$ $\approx$ 4000 $M_{\odot}$.
Such a comparison is necessarily crude, particularly given our assumptions deriving 
the K mag/stellar mass calibration and extending a Salpeter IMF to the extreme lower
main sequence and pre-main sequence.

IRAS~06562$-$0337 appears to be a large and rich young cluster.
Similar young star clusters with massive Ae/Be central stars 
have been the subject of two recent surveys.
Hillenbrand (1995) investigate the inner 0.3 pc of 17
clusters and find a linear relationship between central star
mass and stellar surface density; Testi et al.~(1997) study fields
around 19 Herbig Ae/Be stars and report a correlation between
spectral type and cluster richness, with earlier spectral types
resulting in significantly more cluster stars, and some evidence
of a threshold effect around spectral type B5--B7.  Similar to
Hillenbrand, they find a characteristic cluster radius to be 0.2 pc.
At a distance of 7 kpc (BGG), our 30$\arcsec$ radius 
corresponds to a physical radius of 1 pc.
This is $\sim$5 times the ``typical'' cluster radius,
implying IRAS~06562$-$0337 is a very large cluster.
The smallest distance estimate (2.4 kpc, GMSP)
results in an adopted cluster radius of 0.34 pc, still
$\sim 1.5 - 2$ times the ``typical'' value.  
Barsony et al.~(1991) found 100 cluster
members within $r\sim 1$ pc in the young star cluster LkH$\alpha$ 101,
thus IRAS~06562$-$0337 is not unprecedented.
We find an average stellar surface density 
of $\sim$25/pc$^2$ within 30$\arcsec$, increasing
towards smaller radii.  Testi et al.~define a richness 
indicator, I$_C$,
which measures the stellar density enhancement of a cluster,
independent of cluster size, using K band starcounts.  They find early type stars 
have I$_C > 10$, and MWC 137 (a B0 star) has the
highest value in their sample, of I$_C = 76$.  We estimate
a value of I$_C$ = 103 for IRAS~06562$-$0337.  
The high apparent mass of the
central star is therefore
consistent with the large size and richness
of the IRAS~06562$-$0337 cluster.

\subsection{Comparison of the K$^{\prime}$ Image and CO Map}

We determined the coordinates of the central star
by comparing the positions of 12 stars in our image with the Digitized
Sky Survey image.  The centroid of the central star at K$^{\prime}$ is
offset slightly to the West of the visible image by $\Delta\alpha \sim$0.6$\arcsec$.
Taking this into account, we found the two images to be in excellent
agreement, with RMS scatter of less than 0.5$\arcsec$.  Therefore,
we report the J2000 K$^{\prime}$ (2$\micron$) position
of the central object to be $\alpha$ = 06:58:44.31,
$\delta$ = $-$03:41:09.97 (with uncertainties of $\pm0.5\arcsec$).
This differs by $\Delta\alpha=-$5.7$\arcsec$, $\Delta\delta=-$1.2$\arcsec$ from
the near infrared position given by Manchado et al.~(1989).

Figure 3 presents the CO(2$\rightarrow$1) map
of IRAS~06562$-$0337, graciously
provided by BGG, overlaid on our K$^{\prime}$ image.  The CO data was obtained in 
1993 August and 1997 May with the IRAM 30-m telescope with a beam size of 12$\arcsec$
and an efficiency of 0.45 at 230 GHz.
There is excellent
agreement between the central peak of the CO emission and the
central object in the K$^{\prime}$ image, even better than the agreement
between the IRAS position and the CO emission pointed out by BGG.
The peak in the CO emission and the central star are offset
from each other by
$\Delta\alpha=$0.6$\arcsec$, $\Delta\delta=$3.4$\arcsec$.
The central contour of the CO plot is slightly
elongated, extending 8.4$\arcsec$ E--W and 11.5$\arcsec$ N--S,
which may imply a non-spherical molecular gas distribution.  
The majority of the
emission (the first 12 contours) falls within 30$\arcsec$ of the central star,
lending support to our choice of 30$\arcsec$ for the cluster radius, although
strong emission ($>$ 5 K) extends over a region of $\sim 1\arcmin$ radius.  
The singly-peaked
emission suggests that the central object in the K$^{\prime}$ 
image is the only massive
young star in the cluster.  The coincidence of the cluster stars and nebulosity
seen in our K$^{\prime}$ image with the strong 
CO emission is strong evidence that the cluster
is embedded in its placental molecular cloud (see also discussion in BGG).

The morphology of the molecular cloud around IRAS 06562$-$0337 sets it
apart from other young clusters.  Figure 3 clearly shows the CO distribution
to be roughly spherically symmetric and well centered on the massive central star.
Studies of Herbig Ae/Be clusters still associated with remnant molecular gas
(Hillenbrand 1995, Barsony et al.~1991) do not generally find this to be the case.
In fact, although the sample size is small, in almost every case
the star appears to be located near the edge of the
cloud, which is often elongated or otherwise shows signs of disruption.  This may
suggest interaction between the star and the cloud, such that the energy from
the star is dispersing or disrupting the cloud (Hillenbrand 1995).  We 
speculate that this implies extreme youth in the case of IRAS 06562$-$0337, where
the parent molecular cloud is still in control, keeping the hot,
young star snugly cocooned inside its dense layers.

\section{Optical Spectroscopy}

\subsection{Observations and Data Reduction}

On 1997 April 16 UT, we acquired
3 spectra of IRAS~06562$-$0337 from the Kast spectrograph
(Miller \& Stone 1993) on the 3m telescope 
at Lick Observatory\footnote{We thank A. Filippenko, A. Barth,
and A. Gilbert for kindly obtaining these spectra for us.}.
Spectrum \#1 was exposed on the blue side of Kast for 900 seconds
using the 600 line mm$^{-1}$ grism. 
Spectrum \#2 was exposed concurrently 
on the red side of Kast for 900 seconds using the 
830 line mm$^{-1}$ grating.  Spectrum \#3 was exposed on 
the red side of Kast for 600 seconds, approximately 35 minutes 
after the other spectra, using the 600 line mm$^{-1}$ grating.
All three spectra were taken through a 2 arcsec slit. 
The spectra were reduced and extracted using standard IRAF
routines (Massey, Valdes, \& Barnes 1992).
They were wavelength-calibrated using a HeHgCd arc 
lamp spectrum for the blue spectrum
(\#1), and a HeNeAr arc lamp spectrum 
for the red spectra (\#'s 2 and 3).
Fine corrections to the zero points of the wavelength scales were made via 
inspection of the night sky lines.  The resultant wavelength coverage and 
resolution (mean FWHM of lines in the corresponding arc spectrum) for each 
spectrum are 3095--5206\AA ~(resolution 3.2\AA) for spectrum \#1,
6028--8061\AA ~(resolution 4.2\AA) for spectrum \#2, and
4274--7059\AA ~(resolution 5.5\AA) for spectrum \#3.
The instrumental response
function was removed from the spectra and they were flux-calibrated
using observations of spectrophotometric standard stars obtained
on the same night with the same grating settings; Feige 34
(Massey et al. 1988)  was used for spectrum \#1 and
HD 84937 (Oke \& Gunn 1983) for spectra \#'s 2 and 3.

A composite spectrum of IRAS~06562$-$0337
is shown in Figure 4.  It combines the data from 
3100--5200\AA\ in spectrum \#1,
5200--7050\AA\ in spectrum \#3, and 7050--8050\AA\ in spectrum \#2.
The bottom panel of Figure 4 shows the full wavelength and 
flux scales of the composite spectrum.  The most prominent feature
is a very strong emission line of H$\alpha$ that dwarfs any other
spectral features.  The continuum level increases to the red.
A power law fit to the continuum, 
$F_{\lambda}\propto\lambda^{\alpha}$, gives an index of
$\alpha\approx2.5$.
In the middle panel of Figure 4, we have reduced the range of
the flux axis so that the weaker features are more apparent.
The first three Balmer lines and several terrestrial atmospheric
features are indicated.  We were able to identify Balmer
lines all the way to H11 ($\lambda3771$); 
blueward of H$\delta$, the Balmer lines are in absorption rather than
emission.  Although the spectrum becomes somewhat noisy at its blue end, 
a small central emission component 
(peaking well below the local continuum level) 
is apparent in the bottom of each of the 
H$\epsilon$--H11 absorption lines.  
This suggests that the Balmer profiles in
IRAS~06562$-$0337 are made up of at least two components: 
an absorption line and a superposed emission line.
The absorption depth does not change greatly from H$\epsilon$ to H11,
but the intensity of the emission component 
increases up the Balmer series such that
the lines switch from mostly absorption at H$\epsilon$
to mostly emission at H$\delta$.  This is qualitatively similar
to the Balmer emission observed in other Be stars (Burbidge and Burbidge, 1953).
In the top panel of Figure 4 the wavelength
and flux scales of the composite spectrum are expanded, 
with several typical features labeled.  In addition to 
the Balmer lines and diffuse interstellar bands (DIBs; Herbig 1975), 
we find several absorption lines 
of \ion{He}{1} and many (permitted and forbidden) emission 
lines of metals.  Most of the latter
are lines of \ion{Fe}{2} and [\ion{Fe}{2}]; 
the remainder are \ion{Si}{2} and [\ion{O}{1}].
There are a few unidentified and/or extremely weak features in 
the composite spectrum; 
hence, we cannot rule out the possible presence of lines of other
elements or additional lines of the identified elements.

Tables 1a, 1b, and 1c list the equivalent widths (EWs) and integrated
fluxes for the identifiable lines and DIBs 
in each of the spectra.  
The EWs and fluxes were measured via direct integration 
of pixel values between manually-selected endpoints in the 
flux-calibrated spectra.  Several trials of 
each measurement were accomplished to estimate the uncertainty; 
in general, the measurements are accurate to $\pm10\%$ or better 
(values with larger probable uncertainties are denoted with a ``:''). 
A negative flux value in the tables indicates an absorption line.

\subsection{Spectral Analysis and 1996--1997 Variablity}

The 1996 (KLR) and 1997 (this paper) epoch spectra 
have similar resolution and high S/N, which
allows the first investigation of the variability of many fainter lines.
In Table 2 we have assembled select emission line strengths from our
Tables 1a,b,c and KLR (1996, their Table 1) to make the comparison.
Lines are included in Table 2 if they are relatively unblended
in our 1997 epoch spectra (typical uncertainties of 10\% in the line fluxes)
and were also identified in the KLR spectrum.  In some instances,
particularly useful lines, such as the [\ion{O}{1}] lines, were included
despite their larger flux uncertainties.  
In these cases, we mantain the
use of the ``:'' as in Tables 1a,b,c to designate the larger uncertainty.
In other cases, the 
flux of two blended lines of the same ionization species are listed for
1997, and the combined flux of the two resolved lines in the KLR
spectrum is listed for 1996.
KLR do not report absolute fluxes, so we have
normalized the fluxes to H$\beta$ = 100. 
In the bottom part of Table 2, we have listed several line ratios;
typical uncertainties are 14\% unless otherwise discussed below.

KLR suggest that the variable emission seen by GMSP may be understood
as a patchy circumstellar envelope around a single star.  
Using the Balmer decrements, we investigate
the hypothesis that variable extinction, possibly due to orbiting dust clouds, 
could mimic variable emission between 1996 and 1997.  
We estimate the uncertainties in our flux measurements for 
the H$\alpha$, H$\beta$ and H$\gamma$
Balmer lines is $\approx$ 5\%, or 7\% in the ratios.
KLR do not report uncertainties; however, we will assume
they are 10\%, or 14\% in the ratios.  To derive the extinction,
we adopt the intrinsic decrements H$\alpha$/H$\beta$ = 2.80
and H$\gamma$/H$\beta$ = 0.47 from
Osterbrock's (1974) compilation of Case B recombination line ratios,
appropriate for a temperature near 2$\times$10$^{4}$ K (GMSP), and the
reddening law of Cardelli, Clayton, and Mathis (1987). 
For 1997 we find $A_{V}$ = 8.15 $\pm$ 0.25 and $A_{V}$ = 8.6 $\pm$ 0.4
from H$\alpha$/H$\beta$ and H$\gamma$/H$\beta$ respectively.  Similarly, for 
1996 we find $A_{V}$ = 8.35 $\pm$ 0.4 and $A_{V}$ = 8.2 $\pm$ 0.9.
The weighted average of all four measurements gives
$A_{V}$ = 8.3 $\pm$ 0.35 mag, or a color excess of $E(B-V)$ = 2.68 $\pm$ 0.12 mag.
Note that our estimate of the uncertainty does not include a contribution
from the reddening law or the adopted intrinsic line ratios.
The color excesses derived in 1996 and 1997 agree well, 
which implies a negligible change in extinction over the one year interval.
Our color excess can be compared with that derived by GMSP of $E(B-V)$ = 1.75 $\pm$
0.25 mag.  The discrepancy is significant, and may be due to different assumptions
in the reddening law or intrinsic line ratios or a real change in the
extinction between 1990 and 1996/1997.  GMSP do not give observed
line fluxes, only dereddened fluxes, and do not describe their 
procedure for calculating the reddening in detail. 
For these reasons, we do not make any additional comparisons to the GMSP
spectral line data.  Lastly, the Cardelli, Clayton, and Mathis
reddening law gives $A_{K}$ = 0.114$\times$$A_{V}$, or $A_{K}$ = 0.95 mag
for IRAS~06562$-$0337, which is 
what we assumed in Section 2.

Keenan et al.~(1995) present an [\ion{O}{1}]~(6300+6364/5577)
temperature and density diagnostic diagram, as do Bautista and Pradhan (1995).
In our data, the $\lambda6364$ 
and $\lambda5577$ lines were both blended and have large uncertainties, estimated
to be approximately 30\%.  
The [\ion{O}{1}]~(6300+6364/5577) line ratio (see Table 2) is consistent 
with no change over the 1996 to 1997 interval, 
although this result is fairly uncertain.  
It is noteworthy that the 
unblended and strong [\ion{O}{1}]~$\lambda6300$
line intensity decreased by a factor of two over this year 
(a high significance), yet the [\ion{O}{1}] line ratio remained constant.
We derive log($n_{e}$) $\approx$ 5.8 assuming an electron
temperature, $T_{e}$ = 2$\times$10$^{4}$ K (GMSP).  We additionally derive an upper
limit to the electron density, log($n_{e}$) $\le$ 15.45, from 
the Inglis-Teller formula (1939) and our observation of the H11 Balmer line.
Our use of the Inglis-Teller formula is limited by the
noise at the blue end of our spectra.
For comparison, Viotto (1976) tabulated electron densities and
temperatures for a number of Be stars.  They are typically
$n_{e} \approx 4\times10^{11}$ cm$^{-3}$ and $T_{e} \approx$ 1-2$\times10^{4}$ K,
with earlier spectral types having the higher electron temperatures.
The (low) electron density derived from the [\ion{O}{1}] line ratio implies
an origin for these lines at the ``outskirts'' of the ionized region.  
This may reflect a partially ionized zone at the edge
of a wind-swept ionized cavity (e.g. Barsony et al.~1990, Becker and White, 1988) 
or the ``outer wind'' analagous to that seen in P Cygni (Stahl et al. 1991).

GMSP argued that their time-series of spectral data imply an increasing temperature,
consistent with a rapidly evolving central star in a PPN.
We consider whether the 1996--1997
spectral line variability can be understood as simple changes in electron
temperature or density.
The permitted lines of \ion{Fe}{2} and \ion{Si}{2} show
a general trend of decreasing intensity over the 1996 to 1997 interval.
Of five lines selected for comparison,
one line (\ion{Fe}{2}~$\lambda5018$) shows an $\approx$ 1$\sigma$
increase, while four other lines show 1-2$\sigma$ decreases.  The
average decrease in line intensity is 80\%.
If the permitted lines of \ion{Fe}{2} and \ion{Si}{2} behave similarly
to the hydrogen and helium recombination lines, then their intensities
depend very weakly on electron density (e.g. collisional effects, Osterbrock 1974),
and with a temperature dependence, 
\begin{equation}
I(recombination) \propto T_{e}^{-1.1}
\end{equation}
(Oliva et al., 1989).  This implies an increase in the electron temperature
of approximately 20\% in one year.  Assuming an initial electron temperature
of $2\times10^{4}$ K, the increase would be $\Delta$$T_{e} \approx +4500$ K.
The most dramatically varying line in Table 2 is [\ion{O}{1}]~$\lambda6300$,
which decreased by a factor of two (approximately 5$\sigma$) between 1996 and 1997.
The electron temperature
and density dependence of this forbidden line is given by: 
\begin{equation}
I(forbidden) \propto 
n_{e} n_{ion} \Omega(T) T_{e}^{-1/2} 10^{-T_{\lambda}/T_{e}}
\end{equation}
where $T_{\lambda}$ = c2/$\lambda$, c2 is the second radiation constant
(1.438769 cm K) and $\lambda$ is in units of cm; for [\ion{O}{1}]~$\lambda6300$,
T$_{\lambda}$ = 22837.
$\Omega$(T) is the collision strength which we will assume has a weak
temperature dependence (Osterbrock 1974), 
$n_{e}$ is the electron density, and $n_{ion}$
is the density of the ion species, in this case,
neutral oxygen.  A change in
the electron density of 0.2 dex is reasonably allowed by the 
uncertainty of our [\ion{O}{1}] line ratio.  Therefore, we
cannot rule out a factor of two decrease in the electron density
as the cause of the factor of two decrease in the [\ion{O}{1}]~$\lambda6300$ 
line intensity.
If the flux variability is entirely due to a change in temperature, 
we find $\delta$$T_{e} \approx +9000$ K, with an estimated uncertainty 
of about 5000 K, consistent
with the temperature change implied by the permitted lines.
However, an apparent temperature increase of $\sim$5000 K in one year seems
unlikely to be related to stellar evolution (the suggestion of GMSP, KLR),
and rather implies more complex physical processes. 

The intensity of all lines 
of [\ion{Fe}{2}] listed in Table 2 increased
between 1996 and 1997 and appear anti-correlated with the
intensities of the [\ion{O}{1}] lines, which decreased
in the same interval.
Typically, a positive correlation is
found between [\ion{O}{1}] and [\ion{Fe}{2}] emission strength in a wide range
of objects including supernova remnants, 
Seyfert galaxies, and \ion{H}{2} regions
such as Orion (Bautista and Pradhan, 1997, 1995; Bautista, Pradhan 
and Osterbrock 1994; Mouri et al.~1990).  In these objects, [\ion{O}{1}] emission 
is believed to originate in partially ionized zones (PIZs), lying at
the edges of fully ionized zones.  The [\ion{O}{1}]/[\ion{Fe}{2}]
correlation is attributed to the filling factor of the PIZs and the
coincidence of the [\ion{O}{1}] and [\ion{Fe}{2}] species in the PIZs.
In IRAS~06562$-$0337, a
simple change in the electron temperature or density will not explain 
the behavior of both the [\ion{O}{1}] and [\ion{Fe}{2}] lines.  It is
conceivable that 
dust grains are being blown away and iron depleted onto
the grains is being returned to a gas state, enough to
offset the effect of an increasing temperature. 
An alternative, and preferred, explanation
is that the [\ion{O}{1}] and [\ion{Fe}{2}] emission are not originating
in the same zone, which argues against a presumably static shell-like PIZ at the 
edge of a wind-swept cavity and favors a more dynamic ``outer wind'' origin for
these forbidden lines.  Lastly, we note that the suggestion of KLR that the
[\ion{Fe}{2}] emission is due to flourescence is unlikely at densities of
log($n_{e}$) $\approx$ 5.8, as shown by Bautista, Peng, and
Pradhan (1996).

\section{ Conclusions }

The bright central object in IRAS 06562$-$0337 is probably a single 
Herbig Ae/Be class star, with $M \approx 20 M_{\odot}$.
There is some evidence that it is seen at least partially in reflection.  
For example, the central star is poorly
fit by the stellar point spread function derived from other stars in our
K$^{\prime}$ image.
In addition, the centroids in the visible and at K$^{\prime}$ are offset
by $\Delta\alpha \sim$0.6$\arcsec$.
A comparison of our 1997 epoch optical spectra to the 1996 epoch spectrum 
presented by KLR has provided new evidence for spectral 
line variability in IRAS~06562$-$0337 that is not easily interpreted as 
simple changes in temperature
or density.  This argues against previous suggestions that the 
already-published spectral data reflect
real-time evolutionary changes of the central star.
Understanding the immediate environs of the central star is an
essential first step to better understanding the emission line variability
that has been observed.

We note that our analysis of the 1996--1997 spectral variability of 
IRAS 06562$-$0337 using simple zones and changes in one parameter
(such as the electron temperature or
electron density) is only intended as an interpretive tool and should be 
treated with some caution.  The inconsistencies
of this analysis suggest the complexity of the actual
emission line region.  If the central star is analagous to a Herbig Be star,
then its ``fully ionized zone'' should
have a very high electron density. 
Thus, the low density derived with the [\ion{O}{1}] emission 
implies a stratification of densities, likely in the extended 
atmosphere of the central star in IRAS~06562$-$0337.  
The anti-correlation of [\ion{O}{1}] and [\ion{Fe}{2}]
emission is not easily understood if they arise from a common partially
ionized zone (perhaps located at the edge of a compact \ion{H}{2} region 
around the central star), but is likely consistent with an origin
in a dynamic stratified atmosphere.
Spectral variability is also consistent with a
stellar wind origin.  Stellar winds
from Herbig Be type stars are highly variable, 
extremely complex, and poorly understood (e.g., they are possibly 
modulated by magnetic fields 
or non-radial pulsations; Catala et al. 1993).  
Wind line strengths
and profiles can show nightly variations, some of which are 
correlated with temperature
changes of the star (Scuderi et al. 1994).   The appearance of [\ion{O}{3}]
emission in IRAS 06562$-$0337 during 1990 (GMSP) can be understood as shocks 
in stellar jets from a variable
stellar wind (Hartigan \& Raymond 1993).  
Additionally, we emphasize
that the size of the ionized region in IRAS 06562$-$0337 is small.
Six cm radio observations yielded a surprising non-detection given
the strong H$\beta$ flux, and allow an upper limit angular diameter
of $3.4\times10^{-4}$ arcsec to be set for the ionized region (GMSP).
If we assume the 7 kpc distance estimate of BGG, then the ionized 
region in IRAS 06562$-$0337 has an upper limit diameter of 2--3 AU.
We conclude that
stellar winds in the extended, stratified atmosphere of a very young,
$M \approx 20 M_{\odot}$ star 
is the most plausible explanation for
the puzzling spectral variability of IRAS~06562$-$0337, 
the ``Iron Clad Nebula.''

A valuable time series of spectra of the central object in 
IRAS 06562$-$0337, covering 
a 10 year interval, are now available in the literature.  A useful complementary
study would obtain more densely sampled spectroscopic observations, 
to investigate the short timescale behavior of the emission.  In particular,
correlated line variability and characteristic timescales could be used to
constrain the dimensions of the ionized region, which we have suggested is 
likely to be an extended stellar atmosphere.
We emphasize the need for high resolution and high
S/N spectroscopic observations in the near-UV, and of the entire Balmer series,
as these data can be used to study the extended atmospheres of Be stars
(Burbidge \& Burbidge 1953).  
It is important to build a consistent physical picture of the
environs around the central star in order to answer questions such as 
``is the central star seen in reflection?''
This problem in particular could be addressed with high resolution imaging and
polarization data.  
Lastly, as a young star cluster, IRAS~06562$-$0337 is interesting and notable
for its richness.  Multi-color near-infrared photometric observations 
would allow an accurate measurement of the IMF, and
an investigation into problems such as mass segregation and star formation
efficiencies (e.g., Barsony et al. 1991).

\section { Acknowledgments }

Alves' research at a DOE facility is supported by an Associated Western Universities
Laboratory Graduate Fellowship.  Work at LLNL performed under the auspices of the 
USDOE contract no.~W-7405-ENG-48.  Hoard's research supported by NSF grant 
no.~AST 9217911.  Alves and Hoard thank our respective advisors, Kem Cook and
Paula Szkody.  Rodgers is supported by the NASA Graduate Student Research
Program, and thanks her advisors Bruce Balick and Diane Wooden.  Alves acknowledges
Thor Vandehei for his exuberant assistance while observing at the Lick Observatory.
We also thank Bob Becker, Hien Tran, Carlos de Breuck, and Bill Latter for their
useful comments and discussions that significantly improved this paper.

\clearpage
\begin{deluxetable}{lccl}
\footnotesize
\tablecolumns{4}
\tablewidth{0pt}
\tablecaption{Spectrum \#1}
\tablenum{1a}
\tablehead{
\colhead{$\lambda_{obs}$} &
\colhead{EW} &
\colhead{Flux} &
\colhead{ } \nl
\colhead{(\AA)} &
\colhead{(\AA)} &
\colhead{($\times10^{-15}$ erg s$^{-1}$ cm$^{-2}$)} &
\colhead{Identification}
}
\startdata
3774 & 1.8 & $-$1.6\phs  & \ion{H}{1}~$\lambda3771$ (H11) \nl
3797 & 1.3 & $-$1.2\phs  & \ion{H}{1}~$\lambda3798$ (H10) \nl
3812 & 1.1 &    1.0      & \ion{Fe}{2}~$\lambda3814$ \nl
3837 & 2.0 & $-$2.0\phs  & \ion{H}{1}~$\lambda3835$ (H9) \nl
3856 & 1.4 &    1.4      & \ion{Si}{2}~$\lambda\lambda3854, 3856$ \nl
3892 & 2.1 & $-$2.1\phs  & \ion{He}{1}~$\lambda3889$, 
\ion{H}{1}~$\lambda3889$ (H8)\nl
3916 & \phm{:}0.9:&  \phm{:}0.9: & \ion{Fe}{2}~$\lambda3915$ \nl
3973 & 3.0 & $-$3.0\phs  & \ion{H}{1}~$\lambda3970$ (H$\epsilon$)\nl
4100 & 1.0 &    1.1      & \ion{H}{1}~$\lambda4102$ (H$\delta$) \nl
4170 & \phm{:}0.8:&  \phm{:}1.0: & \ion{Fe}{2}~$\lambda4170$ \nl
4245 & \phm{:}0.9:&  \phm{:}1.1: & [\ion{Fe}{2}]~$\lambda4244$ \nl
4340 & 2.8 &    3.7      & \ion{H}{1}~$\lambda4341$ (H$\gamma$) \nl
4390 & \phm{:}0.7:& \phm{:}$-$1.0:\phs& \ion{He}{1}~$\lambda4388$ \nl
4415 & 0.9 & 1.2  & [\ion{Fe}{2}]~$\lambda4414$,\ion{Fe}{2}~$\lambda4417$ \nl 
4430 & 2.1 & $-$2.9\phs  & DIB 4428\AA \nl
4472 & \phm{:}0.4:& \phm{:}$-$0.6:\phs& \ion{He}{1}~$\lambda4471$ \nl
4494 & \phm{:}0.4:&    \phm{:}0.6:    & \ion{Fe}{2}~$\lambda4491$ \nl
4503 & \phm{:}0.2:& \phm{:}$-$0.3:\phs& DIB 4502\AA \nl
4584 & 0.5        &    0.7            & \ion{Fe}{2}~$\lambda\lambda4584$ \nl
4716 & \phm{:}0.2:& \phm{:}$-$0.4:\phs& \ion{He}{1}~$\lambda4713$ \nl
4731 & \phm{:}0.3:&    \phm{:}0.4:    & \ion{Fe}{2}~$\lambda4731$ \nl
4817 & 0.6   &    1.0            & \ion{Fe}{2}~$\lambda4818$,[\ion{Fe}{2}] \nl
4863 & 18.1\phn   &   31.7\phn   & \ion{H}{1}~$\lambda4861$ (H$\beta$) \nl
4883 & 0.6        & $-$1.1\phs        & DIB 4882\AA \nl
4891 & \phm{:}0.4:&    \phm{:}0.7:    & \ion{Fe}{2}~$\lambda4894$ \nl
4924 & 0.6        &    1.0            & \ion{Fe}{2}~$\lambda4924$ \nl
5020 & 1.0        &    2.0            & \ion{Fe}{2}~$\lambda5019$ \nl
5044 & 0.3        &    0.7            & \ion{Si}{2}~$\lambda5041$ \nl
5058 & \phm{:}0.5:&    \phm{:}1.1:    & \ion{Si}{2}~$\lambda5056$ \nl
5151 & \phm{:}0.3:&    \phm{:}0.6:    & \ion{Fe}{2}~$\lambda5149$ \nl
5160\tablenotemark{a}& \phm{:}1.6:& \phm{:}3.6: & 
[\ion{Fe}{2}]~$\lambda\lambda5158, 5159$\nl
5169\tablenotemark{a}& \phm{:}1.6:& \phm{:}3.6: & 
\ion{Fe}{2}~$\lambda5169$ \nl
\enddata
\tablenotetext{a}{These lines are blended; 
EW and flux measurements are for the combined lines.}
\end{deluxetable}

\clearpage
\begin{deluxetable}{lccl}
\footnotesize
\tablecolumns{4}
\tablewidth{0pt}
\tablecaption{Spectrum \#3}
\tablenum{1b}
\tablehead{
\colhead{$\lambda_{obs}$} &
\colhead{EW} &
\colhead{Flux} &
\colhead{ } \nl
\colhead{(\AA)} &
\colhead{(\AA)} &
\colhead{($\times10^{-15}$ erg s$^{-1}$ cm$^{-2}$)} &
\colhead{Identification}
}
\startdata
4436 & \phm{:}3.4: & \phm{:}$-$4.7:\phs & DIB 4428\AA \nl
4475 & \phm{:}1.4: & \phm{:}$-$1.9:\phs & \ion{He}{1}~$\lambda4471$ \nl
4820 & \phm{:}1.3: & \phm{:} 2.1: & 
\ion{Fe}{2}~$\lambda4818$, [\ion{Fe}{2}] \nl
4868 & 19.0\phn    & 32.0\phn & \ion{H}{1}~$\lambda4861$ (H$\beta$) \nl
5026 & 1.0         &    1.9   & \ion{Fe}{2}~$\lambda5018$ \nl
5164\tablenotemark{a}& \phm{:}1.6: & 
\phm{:}3.5: & [\ion{Fe}{2}]~$\lambda\lambda5158, 5159$\nl
5174\tablenotemark{a}& \phm{:}1.6: & 
\phm{:}3.5: & \ion{Fe}{2}~$\lambda5169$ \nl
5204 & \phm{:}0.4:   &  \phm{:}0.9: & \ion{Fe}{2}~$\lambda5198$ \nl
5267\tablenotemark{a}& \phm{:}1.7:& \phm{:}4.0:& 
[\ion{Fe}{2}]~$\lambda5269$ \nl
5277\tablenotemark{a}& \phm{:}1.7:& \phm{:}4.0:& 
[\ion{Fe}{2}]~$\lambda5278$ \nl
5337 & 0.8             &    2.0             & [\ion{Fe}{2}] \nl
5380 & 0.5             &    1.3             & [\ion{Fe}{2}] \nl
5579 & \phm{:}0.2:     &    \phm{:}0.6:     & [\ion{O}{1}]~$\lambda5577$ \nl
5706 & 0.5             & $-$2.0\phs         & DIB 5705\AA \nl
5750\tablenotemark{a}& \phm{:}0.6:& \phm{:}1.7:& 
[\ion{Fe}{2}]~$\lambda5747$ \nl
5754\tablenotemark{a}& \phm{:}0.6:& \phm{:}1.7:& 
[\ion{Fe}{2}]~$\lambda5754$ \nl
5780 & 1.3             & $-$4.0\phs         & DIB 5778,5780\AA \nl
5797 & \phm{:}0.3:     & \phm{:}$-$0.8:\phs & DIB 5797\AA \nl
5870 & 0.8             &    2.5             & \ion{Fe}{2}$\lambda5865$ \nl
5891\tablenotemark{a}& \phm{:}2.6:& \phm{:}$-$8.4:\phs & 
\ion{Na}{1}~$\lambda5890$ (D-lines) \nl
5897\tablenotemark{a}&\phm{:} 2.6:& \phm{:}$-$8.4:\phs & 
\ion{Na}{1}~$\lambda5896$ (D-lines) \nl
5951\tablenotemark{a}& \phm{:}0.6:& \phm{:}2.1: & 
\ion{Si}{2}~$\lambda5949$ \nl
5959\tablenotemark{a}& \phm{:}0.6:& \phm{:}2.1: & 
\ion{Si}{2}~$\lambda5958$ \nl
5980 & 0.6             &   1.9              & \ion{Si}{2}$\lambda5980$ \nl
6048 & \phm{:}0.4:     &   \phm{:}1.4:      & [\ion{Fe}{2}] \nl
6126 & \phm{:}0.4:     &   \phm{:}1.5:      & \ion{Fe}{2}~$\lambda6130$ \nl
6284 & 3.2             & $-$11.7\phs\phn    & DIB 6284\AA \nl
6302 & 1.1             &    4.1             & [\ion{O}{1}]~$\lambda6300$ \nl
6320 & 0.6             &    2.4             & \ion{Fe}{2} \nl
6348 & 0.6             &    2.2             & \ion{Si}{2}~$\lambda6347$ \nl
6365\tablenotemark{a}& \phm{:}0.9:& \phm{:}3.5: & 
[\ion{O}{1}]~$\lambda6364$ \nl
6374\tablenotemark{a}& \phm{:}0.9:& \phm{:}3.5: & 
\ion{Si}{2}~$\lambda6371$ \nl
6386 & 0.7             &   2.8              & \ion{Fe}{2}~$\lambda6384$ \nl
6443 & \phm{:}0.7:     &   \phm{:}2.9:      & \ion{Fe}{2}~$\lambda6446$ \nl
6564 & 302\phn\phn\phn &1265\phn\phn\phn\phn & 
\ion{H}{1}~$\lambda6563$ (H$\alpha$) \nl
6615 & \phm{:}0.4:     & \phm{:}$-$1.8:\phs & DIB 6614\AA \nl
6659 & \phm{:}1.1:     & \phm{:}$-$4.9:\phs & DIB 6661\AA \nl
6681 & \phm{:}0.4:     & \phm{:}$-$1.8:\phs & \ion{He}{1}~$\lambda6678$ \nl
\enddata
\tablenotetext{a}{These lines are blended; 
EW and flux measurements are for the combined lines.}
\end{deluxetable}

\clearpage
\begin{deluxetable}{lccl}
\footnotesize
\tablecolumns{4}
\tablewidth{0pt}
\tablecaption{Spectrum \#2}
\tablenum{1c}
\tablehead{
\colhead{$\lambda_{obs}$} &
\colhead{EW} &
\colhead{Flux} &
\colhead{ } \nl
\colhead{(\AA)} &
\colhead{(\AA)} &
\colhead{($\times10^{-15}$ erg s$^{-1}$ cm$^{-2}$)} &
\colhead{Identification}
}
\startdata 
6250 & \phm{:}0.2: &       \phm{:}0.5:    & \ion{Fe}{2}~$\lambda6248$ \nl
6286 &        3.0  &  $-$11.8\phs\phn     & DIB 6784\AA \nl
6303 &        1.3  &              4.9     & [\ion{O}{1}]~$\lambda6300$ \nl
6320 & \phm{:}0.6: &       \phm{:}2.2:    & \ion{Fe}{2} \nl
6348 &        0.7  &              2.8     & \ion{Si}{2}~$\lambda6347$ \nl
6366\tablenotemark{a}& \phm{:}0.8: & \phm{:}3.1:& 
[\ion{O}{1}]~$\lambda6364$ \nl
6371\tablenotemark{a}& \phm{:}0.8: & \phm{:}3.1:& 
\ion{Si}{2}~$\lambda6371$ \nl
6388 &        0.8     &              3.3     & \ion{Fe}{2}~$\lambda6384$ \nl
6443 & \phm{:}0.2:    &       \phm{:}0.9:    & \ion{Fe}{2}~$\lambda6446$ \nl
6565 & 283\phn\phn\phn & 1143\phn\phn\phn\phn & 
\ion{H}{1}~$\lambda6563$ (H$\alpha$)\nl
\enddata
\tablenotetext{a}{These lines are blended; 
EW and flux measurements are for the combined lines.}
\end{deluxetable}

\clearpage
\begin{deluxetable}{lcc}
\footnotesize
\tablecolumns{5}
\tablewidth{0pt}
\tablecaption{Comparison of 1996 and 1997 Epoch Spectra}
\tablenum{2}
\tablehead{
\colhead{Line Flux or Flux Ratio} &
\colhead{1997} &
\colhead{1996} 
}
\startdata
\ion{H}{1}~$\lambda4341$ (H$\gamma$)   & 11.5 & 12.2 \nl
\ion{H}{1}~$\lambda4861$ (H$\beta$)    & 100  & 100  \nl
\ion{Fe}{2}~$\lambda5018$             & 5.9  & 5.1  \nl
[\ion{Fe}{2}]~$\lambda5269+\lambda5278$ & 12.5 & 11.9 \nl
[\ion{Fe}{2}]~$\lambda5334$           & 6.3  & 3.2  \nl
[\ion{Fe}{2}]~$\lambda5380$           & 4.1  & 3.1  \nl
[\ion{O}{1}]~$\lambda5577$     & \phm{:}1.9: & 4.5  \nl
[\ion{Fe}{2}]~$\lambda5747+\lambda5754$ & 5.3 & 2.2 \nl
\ion{Si}{2}$\lambda5980$              & 5.9  & 7.1  \nl
[\ion{O}{1}]~$\lambda6300$            & 12.8 & 23.5  \nl
\ion{Fe}{2}~$\lambda6320$           & 7.5  & 12.7 \nl
\ion{Si}{2}~$\lambda6347$             & 6.8  & 8.6  \nl 
\ion{Fe}{2}~$\lambda6384$             & 8.8  & 12.3  \nl
\ion{H}{1}~$\lambda6563$ (H$\alpha$)  & 3953 & 4221  \nl
[\ion{O}{1}]~$\lambda6364$     & \phm{:}5.5: & 6.3  \nl
 & & \nl
H$\alpha$/H$\beta$                   & 39.5 & 42.2 \nl
H$\gamma$/H$\beta$                   & 0.115 & 0.122 \nl
[\ion{O}{1}]~(6300+6364/5577)        & 8.0   & 6.6  \nl
[\ion{O}{1}]~$\lambda6300$/[\ion{Fe}{2}]~$\lambda5334$    & 2.0   & 7.3  \nl
\enddata
\tablenotetext{}{Fluxes normalized so that H$\beta$ = 100;
references: 1997 this paper, 1996 KLR.}
\end{deluxetable}

\clearpage

\begin{figure}
\vskip-1in
\plotone{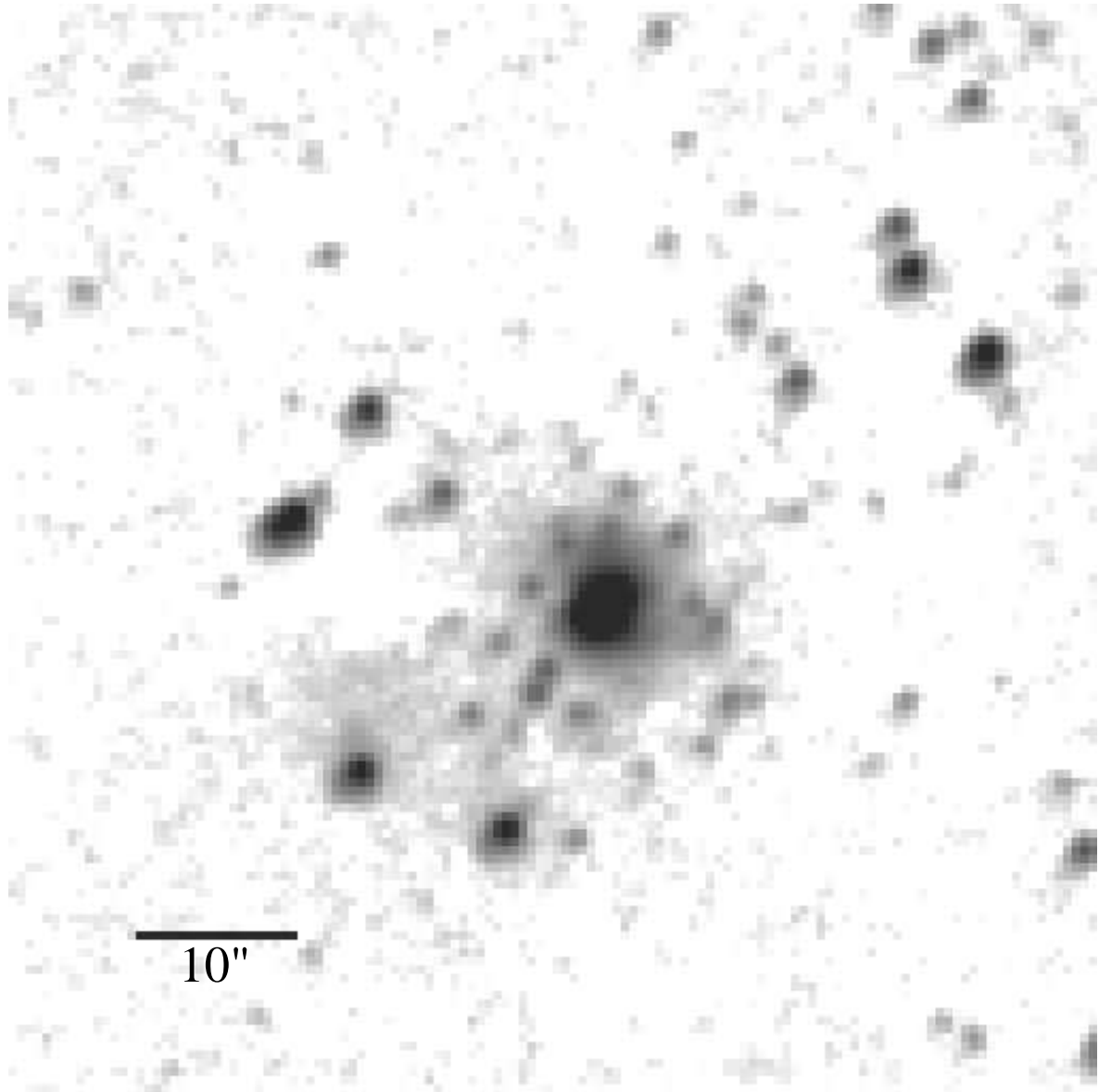}
\caption{Log-scaled K$^{\prime}$ image of IRAS~06562$-$0337 obtained at
Lick Observatory 3m telescope in March, 1997.  Field of view is
70$\arcsec$~$\times$~70$\arcsec$.  North is up, East is to the left. }
\end{figure}

\clearpage

\begin{figure}
\vskip-1in
\plotone{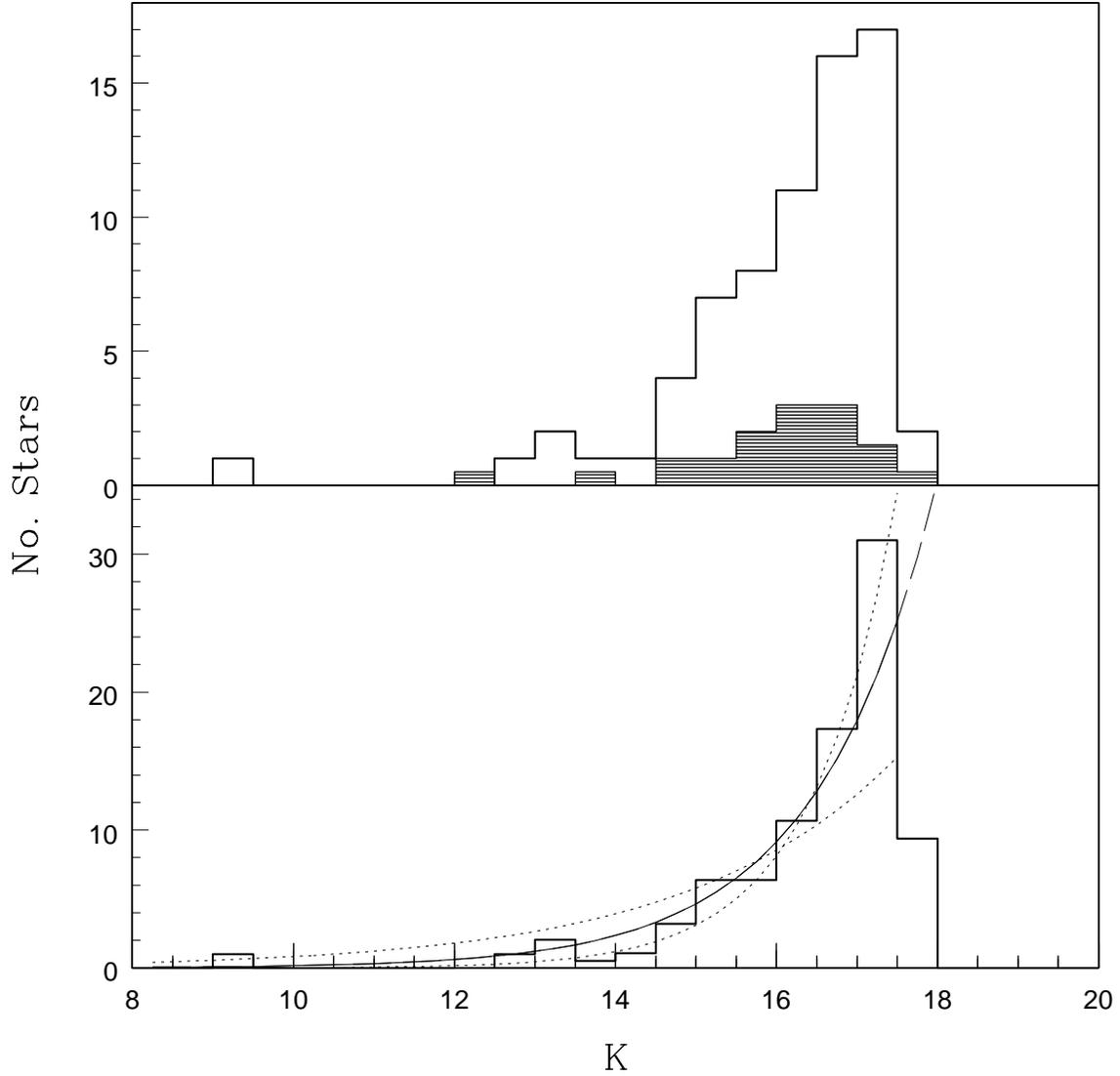}
\caption{Top panel shows K luminosity function of stars inside radius of 30$\arcsec$
around central object of IRAS~06562$-$0337 (71 stars, unshaded histogram) and
average of two equal area control regions (13 stars, shaded histogram).  Completeness
is 90\% at K = 16 mag.  Bottom panel
shows the cluster luminosity function corrected for background/foreground stars and
completeness (histogram) and three model luminosity functions (lines, see text).}
\end{figure}

\clearpage

\begin{figure}
\vskip-1in
\plotone{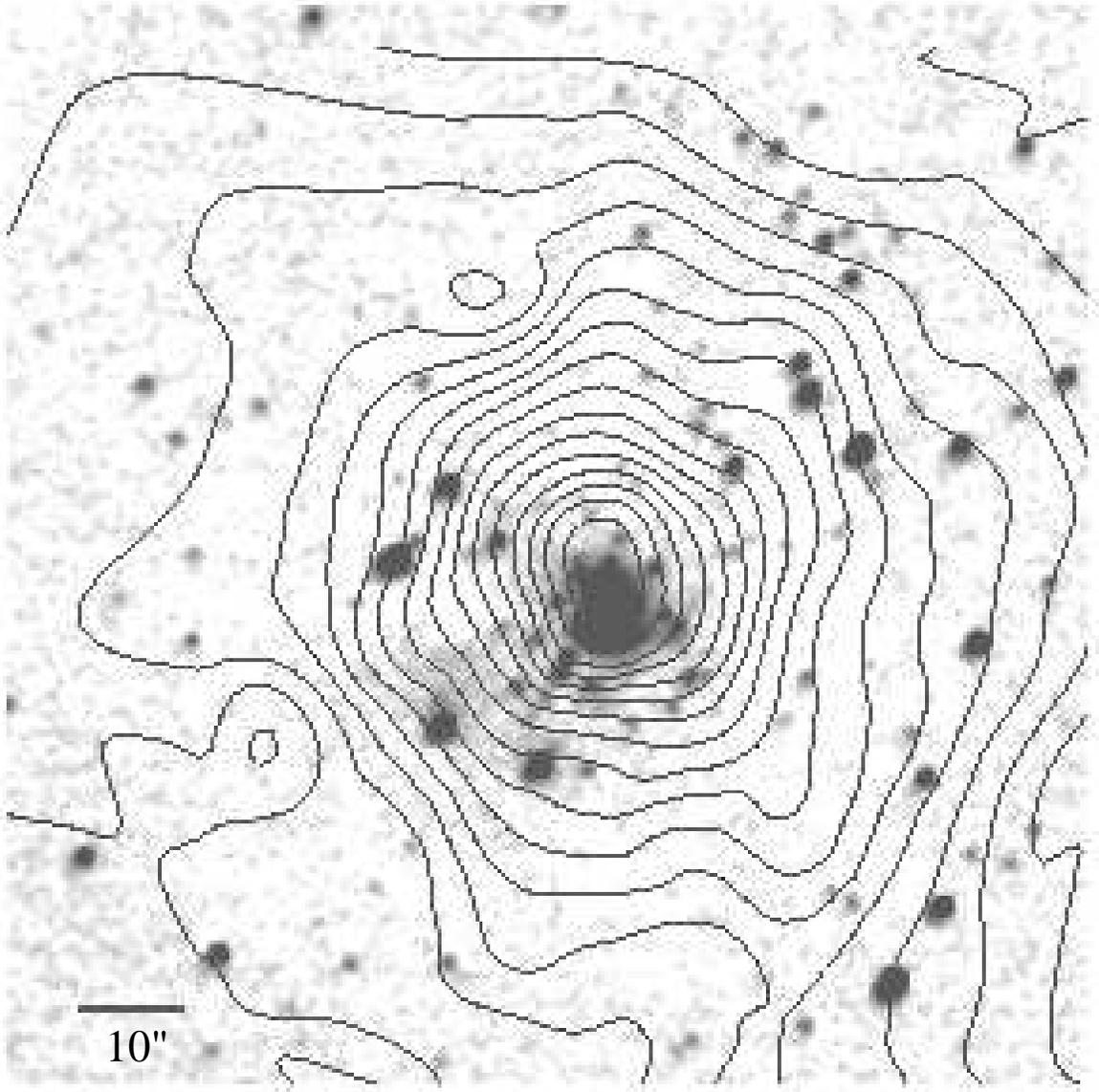}
\caption{Log-scaled K$^{\prime}$ image of IRAS~06562$-$0337 obtained at
Lick Observatory 3m telescope in March, 1997.  Field of view is
approximately 108$\arcsec$~$\times$~108$\arcsec$.  Contours are
CO(2$\rightarrow$1) integrated line intensity from BGG, with the
first contour level and subsequent step size equal to 10 K km s$^{-1}$. }
\end{figure}

\clearpage

\begin{figure}
\vskip-1in
\plotone{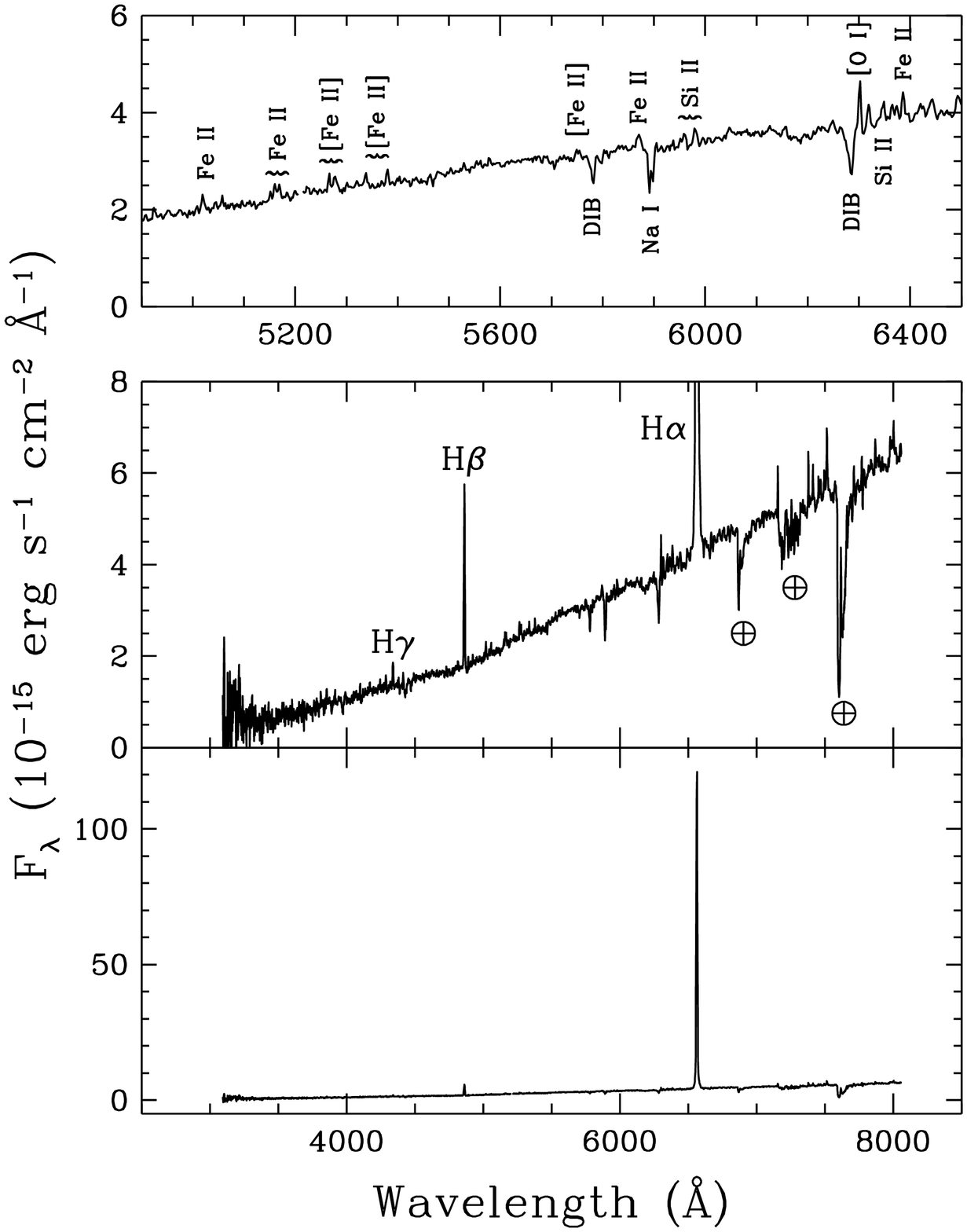}
\caption{The composite optical
spectrum of IRAS 06562$-$0337.  The bottom panel shows the full wavelength and
flux scales of the spectrum.  The middle panel shows the spectrum with an
expanded flux scale.  Several Balmer emission lines and terrestrial
atmospheric absorption features are marked.  The top panel shows a segment of
the spectrum with expanded scales of both wavelength and flux.  A number of
typical features are marked.}
\end{figure}


\clearpage
\begin{references}

\reference{}{ Arribus, S., \& Martinez Rogers, C. 1987, A\&ASS, 70, 303 }

\reference{}{ Bachiller, R, Gutierrez, M.P., \& Garcia-Lario, P. 
1998, A\&A, 311, L45 (BGG)}

\reference{}{ Barsony, M., Schombert, J.M., \& Kis-Halas, K. 1991, ApJ, 379, 221 }

\reference{}{ Barsony, M., Scoville, N.Z., Schombert, J.M., \& Claussen, M.J.
1990, ApJ, 362, 674 }

\reference{}{ Bautista, M.A., Peng, J., \& Pradhan, A.K. 1996, ApJ, 460, 372 }

\reference{}{ Bautista, M.A., \& Pradhan, A.K. 1995 ApJ, 442, L65}

\reference{}{ Bautista, M.A., \& Pradhan, A.K. 1997, astro-ph/9710073 }

\reference{}{ Bautista, M.A., Pradhan, A.K., \& Osterbrock, E. 
1994, ApJ, 432, L135 }

\reference{}{ Becker, R.H., \& White, R.L. 1988, ApJ, 324, 893 }

\reference{}{ Burbidge, G.R., \& Burbidge, E.M. 1953 ApJ, 117, 407}

\reference{}{ Cardelli, J.A., Clayton, G.C., \& Mathis, J.S. 1989, ApJ, 345 245 }

\reference{}{ Carpenter, J.M., Meyer, M.R., Dougados, C., Strom, S.E., \&
Hillenbrand, L.A. 1997, AJ, 114 }

\reference{}{ Catala, C., Bohm, T., Donati, J.F., \& Semel, M. 1993, A\&A, 278, 187}

\reference{}{ Garcia-Lario, P., Manchado, A., Sahu, K.C., \& Pottasch, S.R.
1993, A\&A, 267, L11 (GMSP)}

\reference{}{ Hartigan, P. \& Raymond, J. 1993, ApJ, 409, 705}

\reference{}{ Herbig, G.H. 1960, ApJS, 4, 337}

\reference{}{ Herbig, G.H. 1975, ApJ, 196, 129}

\reference{}{ Hillenbrand, L.A., Strom, S.E., Vrba, F.J., \& Keene, J.
1992, ApJ, 397, 613 }

\reference{}{ Hillenbrand, L.A. 1995, Ph.D. Thesis, Univ. of Massachusetts}

\reference{}{ Inglis, D.R. \& Teller, E. 1939, ApJ, 90, 439}

\reference{}{ Keenan, F.P., Aller, L.H., Hyung, S., \& Brown, P.J.F. 
1995, PASP, 107, 148}

\reference{}{ Kerber, F., Lercher, G., \& Roth, M. 1996, MNRAS, 283, L41 (KLR)}


\reference{}{ Manchado, A., Pottasch, S.R., Garcia-Lario, P., Esteban, C., \&
Mampaso, A. 1989, A\&A, 214, 139 }

\reference{}{ Massey, P., Valdes, F., \& Barnes, J. 1992,
A User's Guide to Reducing Slit Spectra with IRAF,
available on-line http://iraf.noao.edu/ }

\reference{}{ Mihalas, D. \& Binney, J. 1981, Galactic Astronomy (Freeman, New York)} 

\reference{}{ Miller, J. S.\ \& Stone, R. P. S. 1993, Lick Obs. Tech. Rep., No. 66 }

\reference{}{ Misch, A., Gilmore, K., \& Rank, D. 1995, Lick Obs. Tech. Rep., No. 77 }

\reference{}{ Mouri, H., Nishida, M., Taniguchi, Y., \& Kawara, K. 1990, ApJ, 360, 55}

\reference{}{ Oke, J. B.\ \& Gunn, J. E. 1983, ApJ 266, 713 }

\reference{}{ Oliva, E., Moorwood, A.F.M., and Danziger, I.J. 1989, A\&A, 214, 307}

\reference{}{ Osterbrock, D.E. 1974, Astrophysics of Gaseous Nebula and Active
Galactic Nuclei (University Science Books, Mill Valley)}

\reference{}{ Testi, L., Palla, F., Prusti, T., Natta, A., \& Maltagliati, S.
1997, A\&A, 320, 159 }

\reference{}{ Salpeter, E.E. 1955, ApJ, 121, 161 }

\reference{}{ Scuderi, S., Bonanno, G., Spadaro, D., Panagia, N., Lamers, H.,
\& de Koter, A., 1994, ApJ, 437, 465 }

\reference{}{ Stahl, O., Mandel, H., Szeifert, Th., Wolf, B., \& Zhao, F.
1991, A\&A, 244, 467 }

\reference{}{ Stetson, P. 1987, PASP, 99, 191 }

\reference{}{ Viotto, R. 1976, ApJ, 204, 293 }

\end{references}
\end{document}